\documentclass[smallextended]{svjour3}       % onecolumn (second format)

\usepackage{amsbsy}
\usepackage{amsmath}
\usepackage{amssymb}
 \usepackage{float}
\usepackage{booktabs}

\usepackage{fullpage}
\usepackage{bm}
\usepackage{mathrsfs}
\usepackage{enumerate}
\usepackage{multirow}
\usepackage{multicol}
\usepackage{array}
\usepackage{color}
\usepackage{epsfig}
\usepackage{graphicx}	
\usepackage{subfigure}
\usepackage{booktabs}
\usepackage{times}
\usepackage[plain,noend]{algorithm2e}
\usepackage{booktabs}
\usepackage{threeparttable}
\usepackage[titletoc]{appendix}
\usepackage[T1]{fontenc}
\usepackage[utf8]{inputenc}

\usepackage{authblk}
\usepackage{natbib}
\makeatletter
\def\singlespace{\def\baselinestretch{1}\@normalsize}

\renewcommand{\hat}{\widehat}

\begin{document}

\title{Partially Fixed Bayes Additive Regression Trees for  spatial-temporal related model}

\author{Ran Hao         \and
	Bai Yang\thanks{Corresponding author: email@mail.com}
}

\institute{Ran Hao \at
	Department of applied statistics,Shanghai University of Finance and Economics,China \\
	\and
	Bai Yang \at
	Department of applied statistics,Shanghai University of Finance and Economics,China\\}           %  \\

\date{}
\maketitle

\begin{abstract}

Bayes additive regression trees(BART) is a nonparametric regression model which has gained wide
-spread popularity in recent years due to its flexibility and high accuracy of estimation .In spatio-temporal related model,the spatio or temporal variables are playing an important role in the model.The BART models select variables with uniform prior distribution that means treat every variable equally.Applying the BART model directly without properly using these prior information  is not appropriate.This paper is aimed at a modification to the BART by fixing part of the tree's structure.We call this model  partially fixed BART.By this new model we can  improve efficiency of estimation.When we don't know the prior information,we can still use the new model to get more accurate estimation and more structure information for future use.Data experiments and real data examples show the improvement comparing to the original Bart model.

\end{abstract}



\baselineskip=20pt
%%%%%%%%%%%%%%%%%%%%%%%%%%%%%%%%%%%%%%%%%%%%%%%%%%%%%%%%%%%%%%%%%%%%%%%%%%%%%%%%%%%%%%%%%%%%%%%%%%%%%
\section{Introduction}
\label{sec:1}
\indent
\vspace{-5mm}

Bayesian Additive Regression Trees(BART)\citep{chipman2010bart} is a nonparametric regression model that is often more accurate than other tree-based methods such as random forest\citep{breiman2001random},Xgboost\citep{chen2016xgboost}.Compared to other parametric model,it loose some stringent parametric assumptions. It combines the flexibility of a machine learning algorithm with the formality of likelihood-based inference to create a powerful inferential tool.Another advantage for BART model is its robustness to the choice of hyper-parameters.

When we set up new model for real data,we often have some information beforehand that some variables have much inference for the predicted variable.Especially for the spatio-temporal related model.We know that the time or spatio variables play important role in the model.when we know the this part of the model structure,we can set up parametric or semi-parametric model \citep{tan2019bayesian} with the parametric part represent the known structure.But for most situation we don't know the influence structure clearly.We get these information by  logical deduction or business understanding.How can we make full use of this kind of prior information?

Uniform distribution prior is adapted when variables are chosen which means each variable is treated equally.It contradicts with our knowledge that some certain variables are more important than other.One way to facilitate the prior knowledge  is to place more prior to the important variables,but the unequal prior is hard to determined.In this paper We consider fixing the important variables at the root of the trees and we call this model Partially Fixed BART(PFBART).Compared to the original BART model,PFBART can improve the estimation accuracy with right prior knowledge.

The paper proceeds as follows: Section 2 reviews BART, including elements of the MCMC algorithm used for posterior inference. In Section 3 we introduce PFBART in detail.  Some experiments are conducted in section 4 to examine and compare PFBART with BART. Finally, section 5 offers the conclusions of the paper as well as some future works.

\section{Bayesian additive regression trees (BART)}
\label{sec:2}
This section motivates and describes the BART framework.

\subsection{Model}
\label{sec:2.1}

We begin our discussion with the independent continuous outcomes BART because this
is the most natural way to explain BART.For a p-dimensional vector of predictors $X_{i}$ and a response $Y_{i} (1\leq i \leq n)$ ,the BART model posits
\begin{eqnarray}
    Y_{i}=f(X_{i})+\varepsilon_{i},  \varepsilon_{i} \sim N\left(0, \sigma^{2}\right) ,i=1, \cdots, n
\end{eqnarray}
To estimate $f(X)$, a sum of regression trees is specified as

\begin{eqnarray}
    f(X_{i})=\sum_{j=1}^{m} g\left(X_{i} ; T_{j}, M_{j}\right)
\end{eqnarray}
$T_{j}$ is the $j^{th}$ binary tree structure and  $M_{j}=\left\{\mu_{1 j}, \ldots, \mu_{b_{j}}\right\}$is the terminal node parameters associated with $T_{j}$ .$T_{j}$ contains information of which bivariate to split on ,the cutoff value ,as well as the internal node's location.The number of trees $m$ is usually fixed at 200.

 \subsection{Prior }
 \label{sec:2.2}
 The prior distribution for BART model is $P\left(T_{1}, M_{1}, \ldots, T_{m}, M_{m}, \sigma\right)$.Here we assume that
 \\$\left\{\left(T_{1}, M_{1}\right), \ldots,\left(T_{m}, M_{m}\right)\right\}$ are independent with $\sigma$ ,and $\left(T_{1}, M_{1}\right), \ldots,\left(T_{m}, M_{m}\right)$are independent with each other,so we have
  \begin{eqnarray}
  \label{equ:s1}
 \begin{aligned}
P\left(T_{1}, M_{1}, \ldots, T_{m}, M_{m}, \sigma\right) &=P\left(T_{1}, M_{1}, \ldots, T_{m}, M_{m}\right) P(\sigma) \\
& =\left[\prod_{j}^{m} P\left(T_{j}, M_{j}\right)\right] P(\sigma) \\
&=\left[\prod_{j}^{m} P\left(M_{j} \mid T_{j}\right) P\left(T_{j}\right)\right] P(\sigma) \\
&=\left[\prod_{j}^{m}\left\{\prod_{k}^{b_{j}} P\left(\mu_{k j} \mid T_{j}\right)\right\} P\left(T_{j}\right)\right] P(\sigma)
\end{aligned}
  \end{eqnarray}
From $(\ref{equ:s1})$ ,we need to specify the prior for $\mu_{k j} \mid T_{j}, \sigma,$ and  $T_{j}$.
\\
For the convenience of computation, we use the conjugate normal distribution $N\left(\mu_{\mu}, \sigma_{\mu}^{2}\right)$ as  the prior for $\mu_{i j} \mid T_{j}$,$(\mu_{\mu}$,$\sigma_{\mu})$can be derived through computation.
 \\
 The prior for $T_{j}$ is specified by three aspects:
 \begin{itemize}
  \item [1)]
  The probability for a node at depth $d$ to split ,given by $\frac{\alpha}{(1+d)^{\beta}}$ .We can confine the depth of each tree by control the splitting probability so that we can avoid overfitting.Usually $\alpha$ is set to 0.95 and $\beta$ is set to 2.
  \item [2)]
  The distribution on the splitting variable assignments at each interior node,default as  uniform distribution.  Dirichlet distribution are introduced for high dimension variable selection scenario \citep{rockova2017posterior,linero2018bayesianb}.
  \item [3)]
  The distribution for cutoff value assignment,default as uniform distribution.
\end{itemize}

we also use a conjugate prior, here the inverse chi-square distribution for $\sigma$,$\sigma^{2} \sim v \lambda / \chi_{v}^{2}$,the two parameters $\lambda$,$v$ can be roughly derived by calculation.
\subsection{Posterior Distribution}
\label{sec:2.3}

With the settings of priors $(\ref{equ:s1})$,the posterior distribution can be obtained by
\begin{eqnarray}
\label{equ:s2}
\begin{aligned}
P\left[\left(T_{1}, M_{1}\right), \ldots,\left(T_{m}, M_{m}\right), \sigma \mid Y\right] \propto & P\left(Y \mid\left(T_{1}, M_{1}\right), \ldots,\left(T_{m}, M_{m}\right), \sigma\right) \\
& \times P\left(\left(T_{1}, M_{1}\right), \ldots,\left(T_{m}, M_{m}\right), \sigma\right)
\end{aligned}
\end{eqnarray}
$(\ref{equ:s2})$ can be obtained by  Gibbs sampling.First m successive

\begin{eqnarray}
\label{equ:s3}
\begin{aligned}
P\left[\left(T_{j}, M_{j}\right) \mid T_{(j)}, M_{(j)}, Y, \sigma\right]
\end{aligned}
\end{eqnarray}
can be drawn where $T_{(j)}$ and $M_{(j)}$ consist of
all the trees information except the $j^{th}$ tree.Then a draw of $\sigma$ can be obtained from $$I G\left(\frac{\nu+n}{2}, \frac{\nu \lambda+\sum_{i=1}^{n}\left(Y_{i}-\sum_{j=1}^{m} g\left(X_{i}, T_{j}, M_{j}\right)\right)^{2}}{2}\right)$$,$IG$ stands for inverse gamma distribution.
\\

How to draw from $(\ref{equ:s3})$ ? Note that $T_{j}$, $M_{j}$ depends on  $ T_{(j)}, M_{(j)}$ through
$R_{j}=Y-\sum_{w \neq j} g\left(X, T_{w}, M_{w}\right)$
 ,it's equivalent to draw posterior from a single tree of
\begin{eqnarray}
\label{equ:s4}
P\left[\left(T_{j}, M_{j}\right) \mid R_{j}, \sigma\right].
\end{eqnarray}
We can proceed $(\ref{equ:s4})$ in two steps.First we obtain a draw from $P\left(T_{j} \mid R_{j}, \sigma\right)$,then draw posterior from $P\left( M_{j} \mid T_{j},  R_{j}, \sigma\right)$.
In the first step,we have
  \begin{eqnarray} \label{equ:s5}
  P\left(T_{j} \mid R_{j}, \sigma\right) \propto P\left(T_{j}\right) \int P\left(R_{j} \mid M_{j}, T_{j}, \sigma\right) P\left(M_{j} \mid T_{j}, \sigma\right) d M_{j}
  \end{eqnarray}
 ,we call $P\left(R_{j} \mid T_{j}, \sigma\right)= \int P\left(R_{j} \mid M_{j}, T_{j}, \sigma\right) P\left(M_{j} \mid T_{j}, \sigma\right) d M_{j}$ as marginal likelyhood.Because  conjugate Normal prior is employed on $ M_{j}$,we can get an explicit expression of the marginal likelihood.

We generate a candidate tree $T_{j}^{*}$ from the previous tree structure using  MH algorithm.
we accept the new tree structure with probability

  \begin{eqnarray} \label{equ:s6}
  \alpha\left(T_{j}, T_{j}^{*}\right)=\min \left\{1, \frac{q\left(T_{j}^{*}, T_{j}\right)}{q\left(T_{j}, T_{j}^{*}\right)} \frac{P\left(R_{j} \mid X, T_{j}^{*}\right)}{P\left(R_{j} \mid X, T_{j}\right)} \frac{P\left(T_{j}^{*}\right)}{P\left(T_{j}\right)}\right\}.
  \end{eqnarray}
   $q\left(T_{j}, T_{j}^{*}\right)$ is  the probability for the previous tree $T_{j}$ moves to the new tree $T_{j}^{*}$.
The candidate tree is proposed using four type of moves:
 \begin{itemize}
  \item [1)]
Grow,splitting a current leaf into two new leaves,the probability as 0.25.
  \item [2)]
Prune,collapsing adjacent leaves back into a single leaf,the probability as 0.25.
  \item [3)]
Swap,swapping the decision rules assigned to two connected interior nodes,the probability as 0.1.
  \item [4)]
Change,reassigning a decision rule attached to an interior node,the probability as 0.4.
\end{itemize}
Once we have finished sample from  $P\left(T_{j} \mid R_{j}, \sigma\right)$,we can sample the $k^{th}$ tree the $j^{th}$ leaf parameter $\mu_{k j}$ from  $N\left(\frac{\sigma_{\mu}^{2} \sum_{k=1}^{n_{k}} R_{k j}}{n_{k} \sigma_{\mu}^{2}+\sigma^{2}}, \frac{\sigma^{2} \sigma_{\mu}^{2}}{n_{k} \sigma_{\mu}^{2}+\sigma^{2}}\right)$, where $R_{k j}$ is the subset of  $R_{j}$ allocated to the leaf
node with parameter $\mu_{k j}$ and $n_{k}$ is the number of $R_{k j}$ allocated to that node.

\section{Partially Fixed BART}
\label{sec:3}
As mentioned above,uniform distribution is adapted as splitting variables prior which means every variable has equal probability to be chosen.
From reduction or background analysis,we can know that some variables are more important than others in some special models so more probability should be assigned to them such as time variable in the time related model or  location variables in spatial related model.For these situation,directly applying the BART model can't make full use of these prior knowledge.For model $(\ref{equ:s7})$,we run BART model.Figure 1 shows the frequency for each variable to be selected in the model.It can be observed that $x_{1}$ which has global influence is not the highest selected variable, on the contrary some irrelevant variables as $x_{7}$,$x_{9}$ get higher rate than $x_{1}$.
\begin{figure}
\begin{center}
    \includegraphics[scale=0.7]{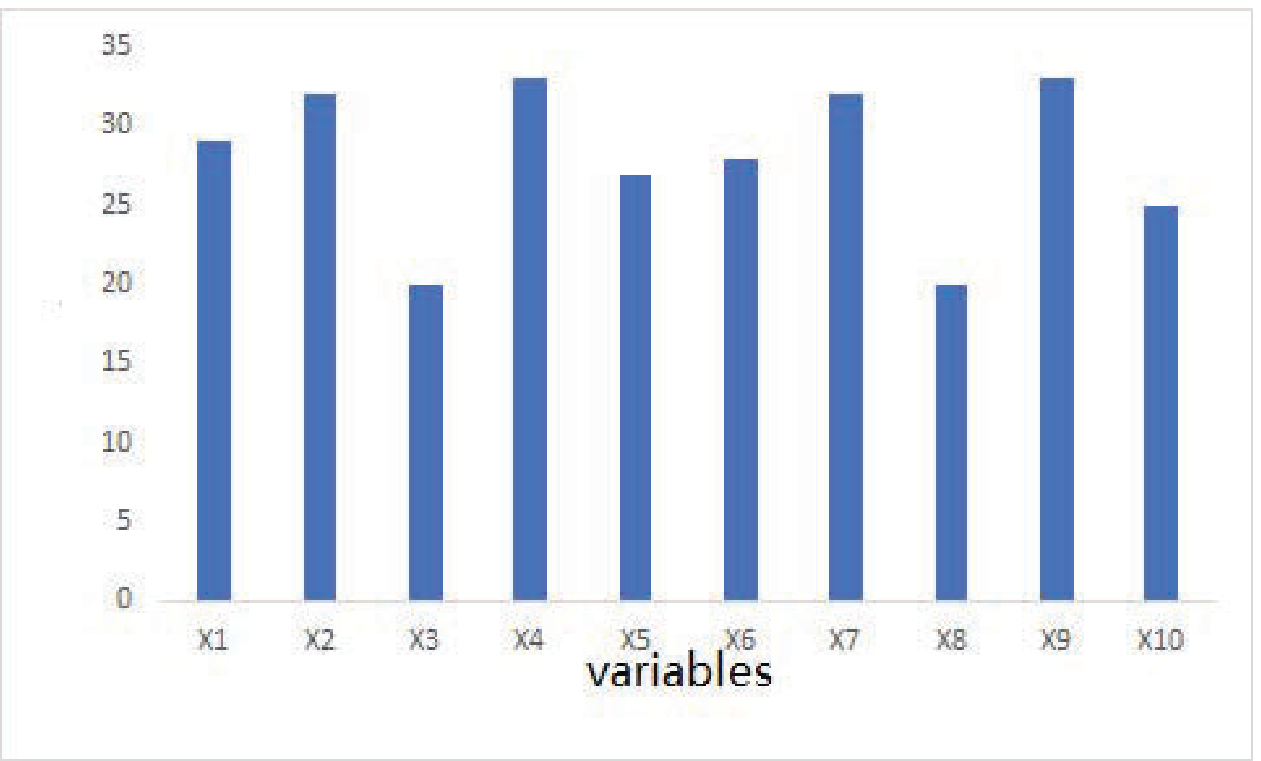}
\end{center}

    \small{\it{{\bf Figure 1.} The frequency of each variable used in the BART model,$X_{1}$ has global influence ,$X_{6},\ldots,X_{10}$ are irrelevant variables for the model.  }}
\end{figure}

When we have such prior knowledge ,we can fix these variables to the bottom of trees.Note that,when the splitting variable $x$ is ordinal ,samples go to the left child node are these with value $x \leq c $ ,c is the cut point for splitting variable,samples with   value $x > c $ go to right child.When multiple layers of variables need to be fixed,it is not unrealistic to assign left child or right child with different splitting variables,so we fix the variables with layers.For example,we know two important variables,we can fix these two variables to the top two levels of the trees,and other variables have no chance to appear at the two levels of the trees.
So we should pay attention to the four steps in generating new tree structure:
\begin {itemize}
  \item [1)]
Grow if the node to grow is in the fixed layers,only the assigned important variables are allow to be chosen as splitting variables.
  \item [2)]
Prune at this stage we will add a logic hyper parameter $pr$.If $pr$ is false and the node to be pruned is in the fixed layers,this process will be terminated.
  \item [3)]
Swap if the two nodes chosen to swap violate the fixed rule,this process will be terminated.

  \item [4)]
Change if the node to be changed is in the fixed layer ,the variable is confined to the fixed variable scope.
\end{itemize}
Here we introduce a logical parameter $swap$ to the model when we have more than one important variables to fixed.$Swap$ is true means that these variables can appears at any fixed layer .$Swap$ is false stands for that the variables to be fixed are in order,in another word,the first important variable is the only variables can be selected in the first layer of the trees.
Considering that BART model limit the trees from growing too deep,if we fix too many variables at the top of the trees,and then it's hard for other variables to be included in the trees.So we introduce a logical parameter $cp$.We keep the splitting probability unchanged when $cp$ is false.When $cp$ is true, Nodes at the fixed layers have the same splitting probability as the root node of the trees,for nodes not in the fixed layers,the splitting probability is changed to $\alpha (1+d-h)^{-\beta}$ where the $h$ stands for the height of the fixed layers.

\section{Application}
\label{sec:4}

\subsection{Application}
\label{sec:4.1}
\begin{figure}
\begin{center}
    \includegraphics[width=0.8\textwidth]{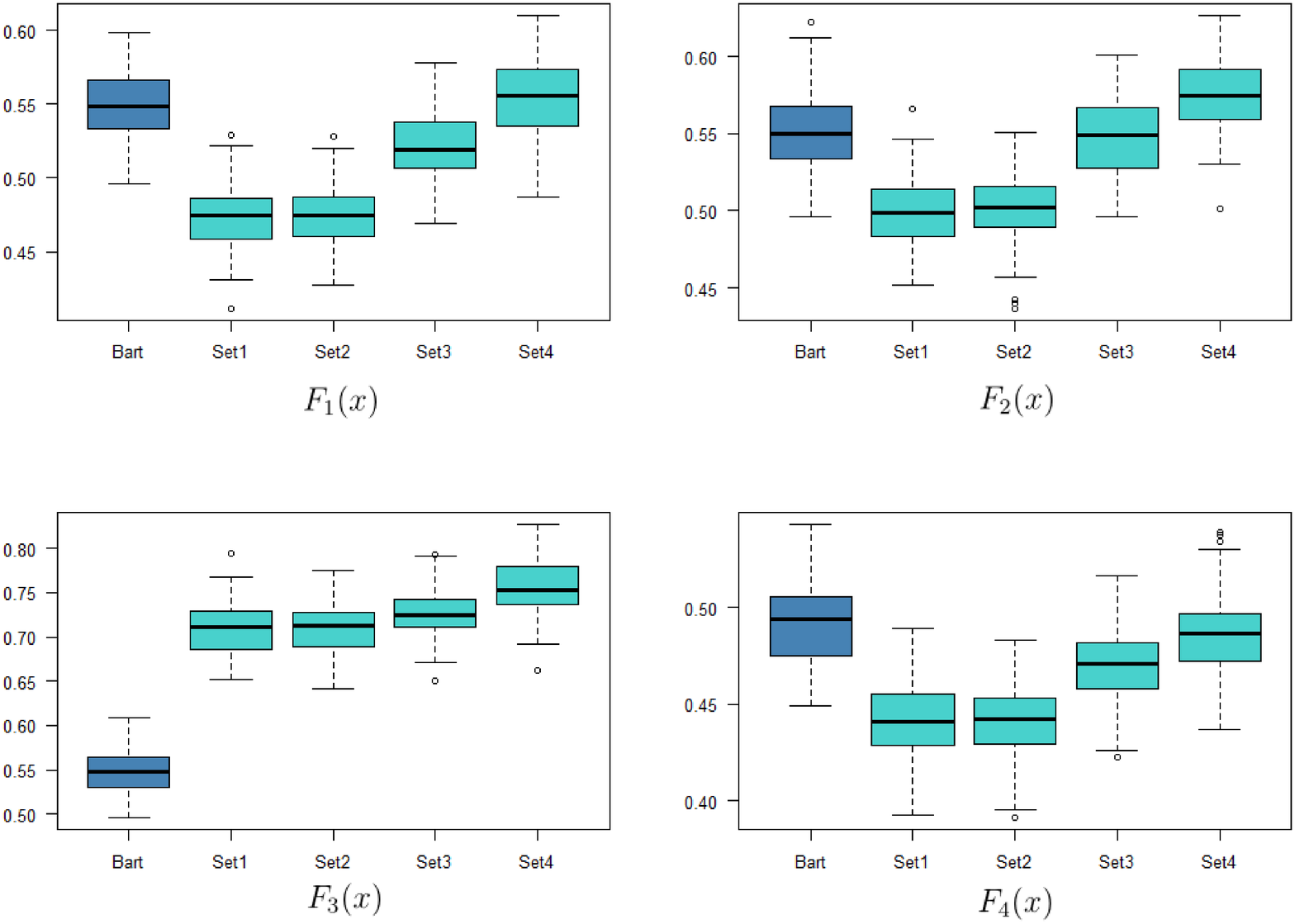}
\end{center}
    \small{\it{{\bf Figure 2.} Boxplots of the RMSE values for each method across the 100 data sets .}}
\end{figure}
 We first proceed to illustrate the benefit of PFBART compared to BART.We  generate data according to the functions below

  \begin{eqnarray} \label{equ:s7}
F_{1}(x)=10sin(\pi X_{1}X_{2} )+5 X_{1}^{2} (X_{3}-0.5) +10 X_{1}^{3}  X_{3} X_{4} +5 X_{1}^{4}X_{5}
  \end{eqnarray}
To make comparation,considering another two functions ,

  \begin{eqnarray} \label{equ:s8}
F_{2}(x)=10sin(\pi X_{1}X_{2} )+5 X_{2}^{2} (X_{3}-0.5) +10 X_{1}^{3}  X_{3} X_{4} +5 X_{1}^{4}X_{5}
  \end{eqnarray}
  \begin{eqnarray} \label{equ:s9}
F_{3}(x)=10sin(\pi X_{6}X_{2} )+5 X_{6}^{2} (X_{3}-0.5) +10 X_{6}^{3}  X_{3} X_{4} +5 X_{6}^{4}X_{5}
  \end{eqnarray}
In $(\ref{equ:s7})$,$X_{1}$ has global influence.In $(\ref{equ:s8})$, the second part have nothing to do with $X_{1}$.In $(\ref{equ:s9})$,$X_{1}$ is an irrelevant variable to the model.

For each function, We generate$100$ data sets in which the sample size is $4000$.
Each data set contains $10$ variables,$X_{1},\cdots,X_{10}$ are random sample from uniform distribution $U(0,1)$.
Every data set is equally divided into two parts,one is used as training,another is left for test.

For both BART and PFBART,  500 burn-in steps and 1000 iterations were used in the MCMC part.Other parameters  are using the default setting.
To prove that the effect of PFBART is not because of the process of variable selection.Based on $(\ref{equ:s7})$,we run the model only with  $X_{1},\cdots,X_{5}$.We call  this setting $F_{4}(x)$.

Based on each training set,each function was then used to predict the corresponding test set and evaluated on
the basis of its  $\mathrm{RMSE}=\sqrt{\frac{1}{2000} \sum_{i=1}^{2000}\left(\hat{f}\left(x_{i}\right)-f\left(x_{i}\right)\right)^{2}}$.

Boxplots of the 100 RMSE values for each function are shown in Figure 2.We generate four setting according the logical parameter mentioned above.
\begin{table}
\begin{center}
\caption{Four settings of the logical parameter.}
\begin{tabular}{|c|c|c|}
\hline
     & Change Prior & Prune       \\
\hline
SET1 & Not Change & Allow \\
\hline
SET2 & Not Change & Not allow   \\
\hline
SET3 & Change & Allow  \\
\hline
SET4 & Change & Not allow  \\
\hline
\end{tabular}
\end{center}
\end{table}

From figure 2,we can see that the model without changing the splitting probability will get more accurate estimation.There is nearly no difference between Set1 and Set2.So next we mainly discuss the comparison between PFBART set1 and BART.For $F_{1}$,PFBART reduce the median of the RMSE by about $15\%$. For  $F_{2}$ that part of the model is irrelevant to the assigned important variable,PFBART reduce the median of RMSE by about $9\%$.When the assigned important
variable has nothing to do with the model,$F_{3}$,we get countereffect when PFBART is applied.If we drop off all the irrelevant variables,we can show that ignoring the effect of variable selection PFBART can overperform BART by $10\%$.

\subsection{UCI Data Sets}
\label{sec:4.2}
When processing data without any prior knowledge,we can still try to fix each variable at the top of the trees and compare the effect of PFBART with BART.We can use these measures of importance to get more information about the structure of the model for further use.
\begin{table}
[h]
\caption{UCI Data Sets}
\label{sect2}
\begin{center}
\begin{tabular}{|c|c|c|}
\hline
Data Set Name                 & Size  & covariate \\ \hline
Tecator                       & 240  & 13    \\ \hline
Abalone                       & 4170 & 8     \\ \hline
Concrete Compressive Strength & 1030 & 8     \\ \hline
Forest Fire                   & 510  & 12    \\ \hline
wine quality                  & 4890 & 11    \\ \hline
Yacht Hydrodynamics           & 300  & 6     \\ \hline
Air Foil Self Noise              & 1500 & 5     \\ \hline
Strike                        & 620  & 6     \\ \hline
Bike Rental                   & 730  & 10    \\ \hline
QSAR aquatic toxicity         & 540  & 8     \\ \hline
QSAR fish toxicity            & 900  & 6     \\ \hline
\end{tabular}
\end{center}
\end{table}
\\Now we select 10 data sets from UCI data sets\citep{Dua2019}.The detail of the data sets can refer to $Table \ref{sect2}$.

To simplification,we randomly delete samples from the data set so the total sample can be  divide exactly by 10.For each data set,we use 10-fold cross-validation.We randomize each data set by 10 times.Here we use the relative RMSE which we defined as the PFBART RMSE divided by BART RMSE for the same data set.So for every covariate we obtain 10 such statistics then we can get figure 3.For PFBART,we choose previously defined Setting SET1.

\begin{figure}
\begin{center}
    \includegraphics[width=0.95\textwidth]{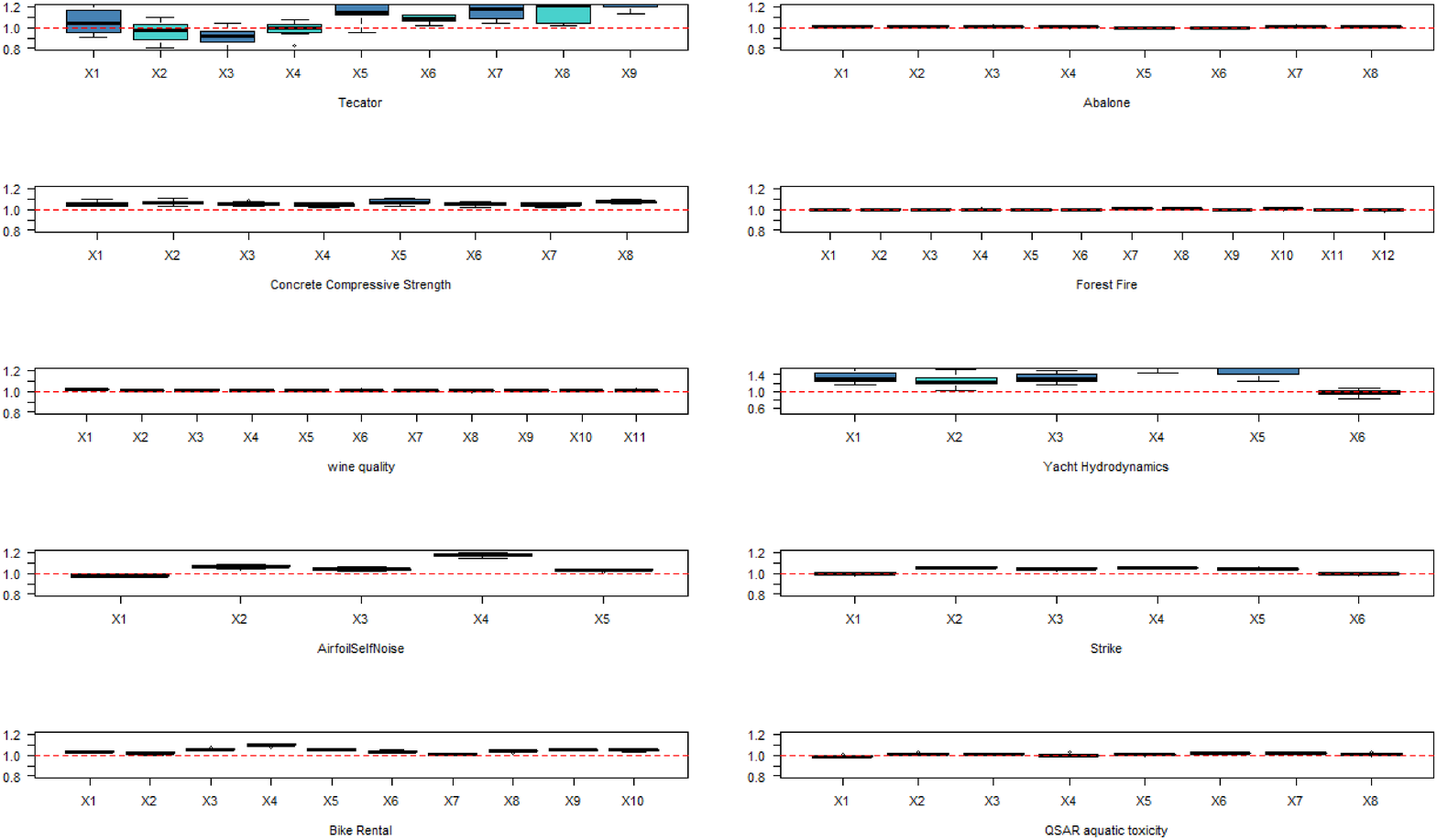}
\end{center}
    \small{\it{{\bf Figure 3.} Relative RMSE for every covariate  in the ten data sets.}}
\end{figure}
In Tecator data,we can see that fixing $X_{3}$ can improve the estimation accuracy that the median of the relative RMSE reaches $91\%$.
\\In Yacht Hydrodynamics,fixing $X_{6}$ will improve a little,same as $X_{1}$ in Air Foil Self Noise.These information may help when constructing model.
\\For Abalone,Forest Fire,Wine quality,Strike and QSAR aquatic toxicity,fixing every covariate has almost the same effect with BART model.That means these covariates are contributing to the model and are not irrelevant to the model,for another these models are not model with one global influence covariate.

For Concrete Compressive Strength,fix each covariate   will lead to worse estimation result,but we found these  covariates  are not irrelevant variables,so we can combine these information with  the background knowledge for later use.

For  Yacht Hydrodynamics,fix covariate except $X_{6}$ will lead to worse estimation specially for  $X_{4}$,but deleting it from model will also lead to worse estimation,that means $X_{4}$ should be included in the model and it is not a variable with global influence,the same is $X_{4}$ of  Air Foil Self Noise and
$X_{4}$ of Bike Rental.

\subsection{ Beijing Housing Price}
\label{sec:4.3}
\begin{table}[h]\label{table2}
\caption[Combination of the parameter]
{Combination of the parameter}
\begin{center}
\begin{tabular}{|c|c|c|c|}
\hline
     & Change Prior   & Prune     & Swap  \\ \hline
SET1 & Change & Allow & Not Allow \\ \hline
SET2 & Change & Allow & Allow  \\ \hline
SET3 & Not Change & Not Allow & Not Allow \\ \hline
SET4 & Not Change & Not Allow & Allow  \\ \hline
SET5 & Change & Not Allow & Not Allow \\ \hline
SET6 & Change & Not Allow & Allow  \\ \hline
SET7 & Not Change & Allow & Not Allow \\ \hline
SET8 & Not Change & Allow & Allow  \\ \hline
\end{tabular}
\end{center}
\end{table}
we use the Beijing house price data \citep{Kaggle2018} to demonstrate the effect of fixing multiple variables in spatial-temporal related model.We use the unit house price as response variable on covariates of location, floor, number of living rooms and bathrooms, whether the unit has an elevator and  some other variables.From our knowledge of real estate,the variables of location and year of trading have great influence to the model.Here we fix the longitude ,latitude and  year of trading at the top three level of the trees.

After preprocess,the total sample size is 296255.Considering the sample size is too large and MCMC iterations is time consuming.We randomly select $30\%$ of the total sample as training samples and keep the $70\%$ for test,by this means we generate 10 data sets.For each data set,we run PFBART with the 8 settings of the combination of the logical parameters listed in Table 3.Here we use the relative RMSE as evaluation statistics.We get the result of Figure 4.

\begin{figure}
\begin{center}
    \includegraphics[width=0.8\textwidth]{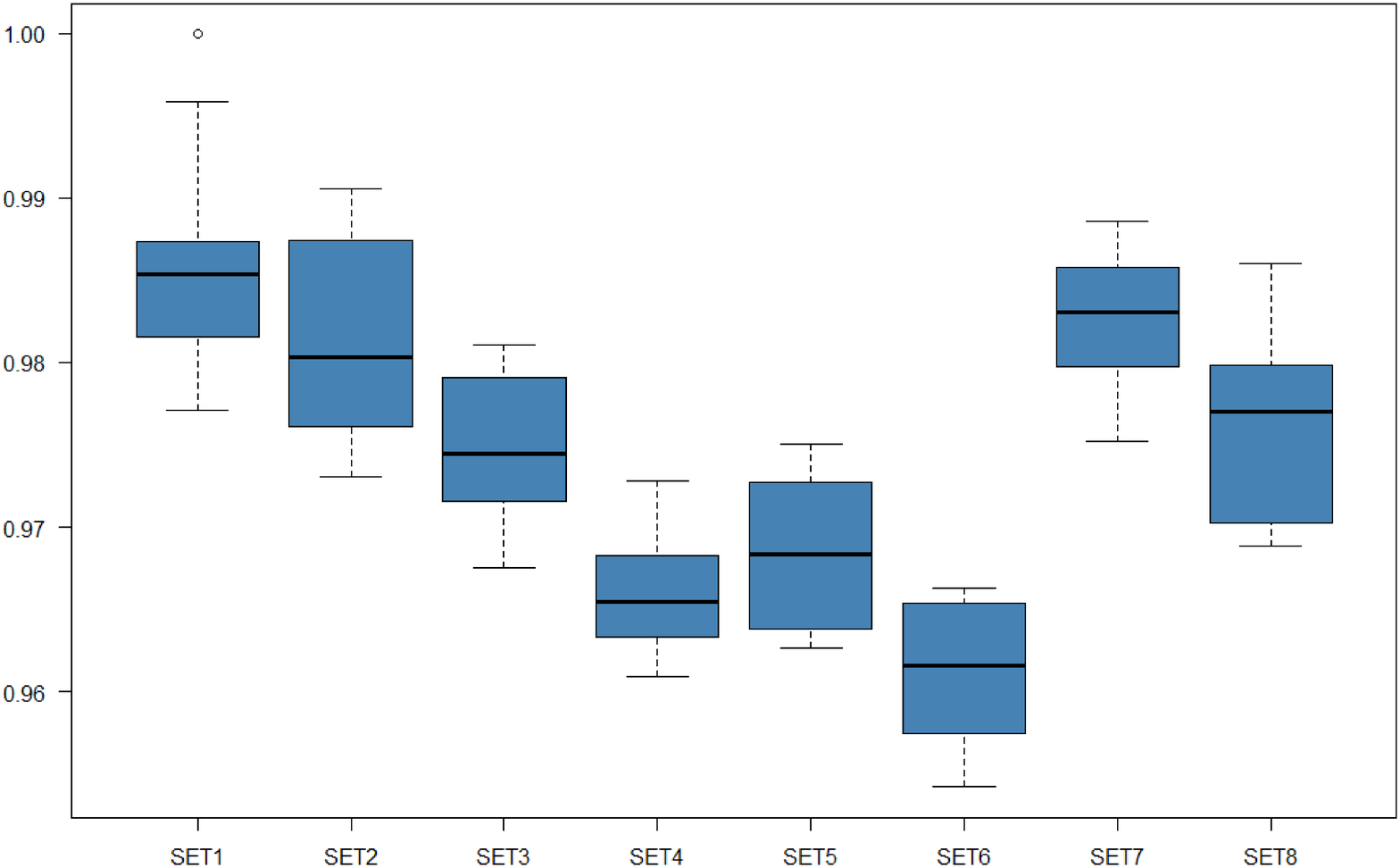}
\end{center}
    \small{\it{{\bf Figure 4.} PFBART relative RMSE for Beijing house price data.}}
\end{figure}
From Figure 4 we can see that for eight combination of PFBART will all improve the estimation more or less in which the set 6 is the best.So we can say that

  \begin{itemize}
  \item [1]
Fixing many layers to the tree has the side effect that other covariate is hard to include into the model.By changing the splitting probability we can solve the problem and treat the  non Fixed layers as if they grow from the root.

  \item [2]

When many layers need to be fixed,another option to more easily include the non fixed part is to prevent nodes to prune at the fixed level.
  \item [3]
When we have more than one layers to fix,we often don't know the order of the importance.Loosing this restriction the model is  more flexible to be a good approximation to the true model.

\end{itemize}

\section{Conclusion and Looking Forward}
\label{sec:5}
When we are building statistics model,especially for spatial-temporal related model,we can know that some variables have global influence to the model by logical deduction or background analysis.This paper demonstrate  a method to make use of these prior knowledge by fixing these important variables to the top of the regression trees.We call this model Partially Fixed BART.Data experiments and real data examples show that improvement can be achieved comparing to the original Bart model.If we don't know the prior information beforehand,we can still use the new model to get more accurate estimation or use the relative statistics as variable importance.\\
The major work of this paper is to develop BART to PFBART.\citet{linero2018bayesianb}introduced a soft BART model which is better suited to approximating   continuous or differentiable functions.In the future We are planning to fix important variables based on the soft BART model and to see whether improvement can be made by this modification.
One of the BART model's advantage is that it can detect interactions between different covariates in the model,maybe it is possible for us to deduct the relationship between covariates and  build a more accurate model based on the structure information we got.
\vskip20pt
\def\refhg{\hangindent=20pt\hangafter=1}
\def\refmark{\par\vskip 1mm\noindent\refhg}

\bibliographystyle{spbasic}

\bibliography{Partial2}

\begin{thebibliography}{8}
\providecommand{\natexlab}[1]{#1}
\providecommand{\url}[1]{{#1}}
\providecommand{\urlprefix}{URL }
\expandafter\ifx\csname urlstyle\endcsname\relax
  \providecommand{\doi}[1]{DOI~\discretionary{}{}{}#1}\else
  \providecommand{\doi}{DOI~\discretionary{}{}{}\begingroup
  \urlstyle{rm}\Url}\fi
\providecommand{\eprint}[2][]{\url{#2}}

\bibitem[{Breiman(2001)}]{breiman2001random}
Breiman L (2001) Random forests. Machine learning 45(1):5--32,
  \doi{https://doi.org/10.1023/A:1010933404324}

\bibitem[{Chen and Guestrin(2016)}]{chen2016xgboost}
Chen T, Guestrin C (2016) Xgboost: A scalable tree boosting system. In:
  Proceedings of the 22nd acm sigkdd international conference on knowledge
  discovery and data mining, pp 785--794,
  \doi{https://doi.org/10.1145/2939672.2939785}

\bibitem[{Chipman et~al.(2010)Chipman, George, McCulloch
  et~al.}]{chipman2010bart}
Chipman HA, George EI, McCulloch RE, et~al. (2010) Bart: Bayesian additive
  regression trees. The Annals of Applied Statistics 4(1):266--298,
  \doi{https://doi.org/10.1214/09-AOAS285}

\bibitem[{Dua and Graff(2017)}]{Dua2019}
Dua D, Graff C (2017) {UCI} machine learning repository.
  \urlprefix\url{http://archive.ics.uci.edu/ml}

\bibitem[{Kaggle(2018)}]{Kaggle2018}
Kaggle (2018) Housing price in beijing.
  \urlprefix\url{https://www.kaggle.com/ruiqurm/lianjia/home}

\bibitem[{Linero and Yang(2018)}]{linero2018bayesianb}
Linero AR, Yang Y (2018) Bayesian regression tree ensembles that adapt to
  smoothness and sparsity. Journal of the Royal Statistical Society: Series B
  (Statistical Methodology) 80(5):1087--1110,
  \doi{https://doi.org/10.1111/rssb.12293}

\bibitem[{Rockov{\'a} and van~der Pas(2017)}]{rockova2017posterior}
Rockov{\'a} V, van~der Pas S (2017) Posterior concentration for bayesian
  regression trees and forests. arXiv preprint arXiv:170808734

\bibitem[{Tan and Roy(2019)}]{tan2019bayesian}
Tan YV, Roy J (2019) Bayesian additive regression trees and the general bart
  model. Statistics in medicine 38(25):5048--5069,
  \doi{https://doi.org/10.1002/sim.8347}

\end{thebibliography}

\end{document}